\documentclass[a4paper,10pt,twoside]{cpc-hepnp}

\usepackage{multicol}
\usepackage{graphicx}
\usepackage{booktabs}
\usepackage{amssymb,bm,mathrsfs,bbm,amscd}
\usepackage[tbtags]{amsmath}
\usepackage{lastpage}
\usepackage{CJK}
\usepackage{natbib}

\begin{document}

\begin{CJK*}{GBK}{song}



\title{A general detector testing system using cosmic rays \thanks{Supported by NSFC (11075095) and Shandong Province Science Foundation (ZR2010AM015) }}

\author{%
      XU Tongye$^{1}$
\quad DU Yanyan$^{1}$
\quad SHAO Ruobin$^{1}$\\
\quad WANG Xu$^{1}$
\quad ZHU Chengguang$^{1;1)}$\email{zhucg@sdu.edu.cn}%
}
\maketitle

\address{%
$^1$ MOE key lab on particle physics and particle irradiation,\\Shandong University, Ji'nan 250100, China
}

\begin{abstract}
A cosmic ray hodoscope with two-dimensional spacial sensitivity and good time resolution has been developed. The system is designed to use the cosmic muons as probes to test the performances of charged particle sensitive detectors. This paper will present the structure of this system, the timing calibration and the resulted performance of this system. The results of the test of the prototype electron detector for LHAASO project are presented as well.
\end{abstract}




\begin{multicols}{2}
\section{Introduction}
A COmic RAy Reference System (CORARS) making use of the cosmic muons as radioactive source was developed to study the performances of charged particle sensitive detectors. The system provides the information of arriving time and position of the cosmic muons, which are used to be compared with the measurement of tested detectors, to find out the performance of the tested detectors, such as, time resolution, detection efficiency and so on.

The Large High Altitude Atmosphere Shower Observation (LHAASO)~\citep{Cao2010} is about to be constructed in Yunnan provence, China. An array consists of 5137 electron detectors (ED) is an essential part of this experiment. To test the EDs' time resolution, detection efficiency, and photoyield of single particle is the first application after CORARS was built. Each ED consists of 16 $25cm\times25cm$ scintillator blocks, with $16\times8$ light fibers to collect light to one PMT.

In section 2 we describe the detector system, the electronics setup, and the DAQ (Data Acquisition) of CORARS. Section 3 is devoted to its implementation and obtained resolution . Results of the first testing of ED prototype are given in section 4, followed by a summary.

\section{Description of CORARS}

\label{}
\subsection{The detector system}
\begin{figure*}[!hbt]
  \begin{minipage}{0.45\textwidth}
  \centering
    \includegraphics[width=0.8\columnwidth,height=8cm, viewport=0 0 490 800,clip,scale=0.4]{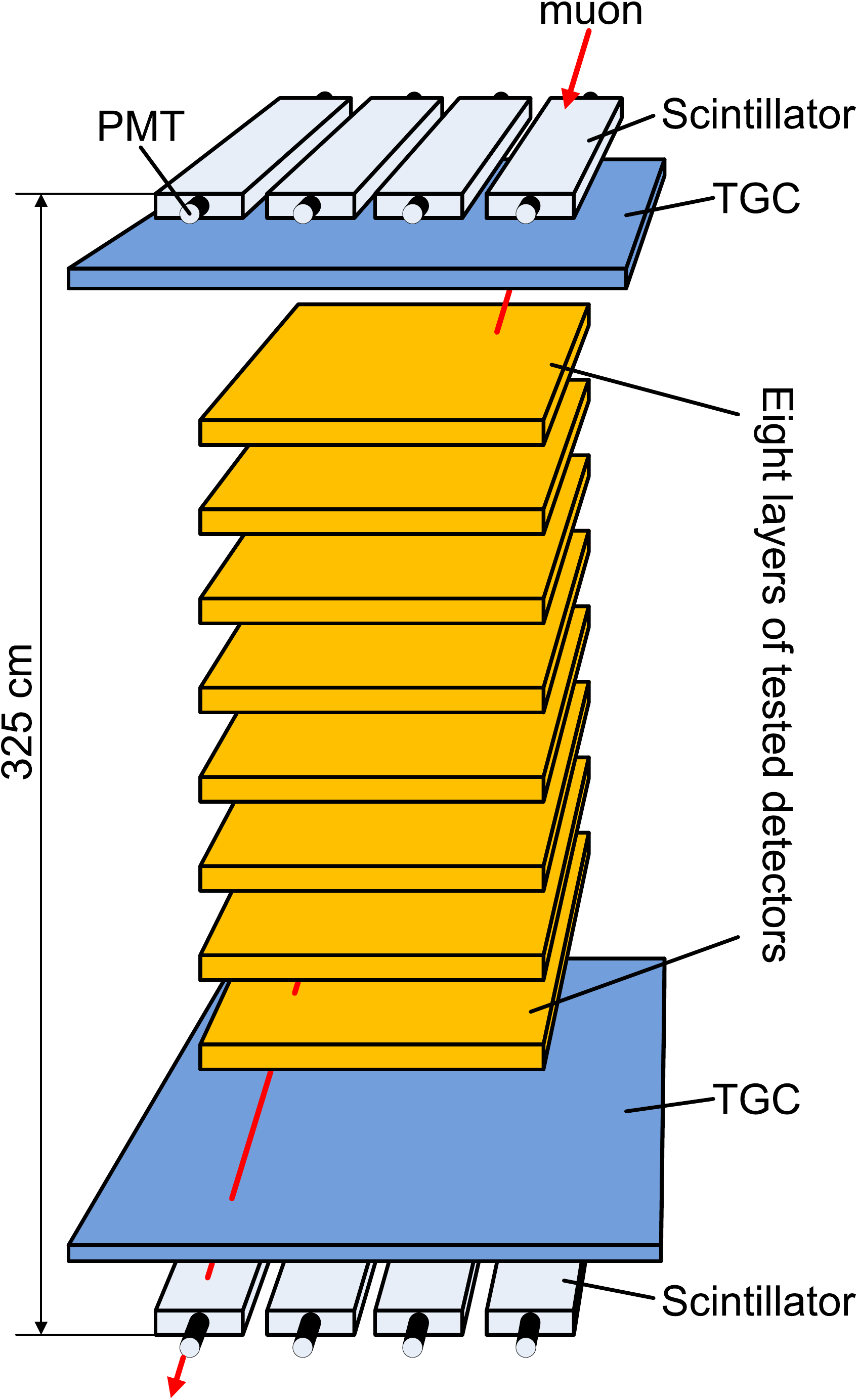}
  \end{minipage}
  \hspace{3ex}
  \begin{minipage}{0.45\textwidth}
    \centering
    \includegraphics[width=0.6\textwidth,viewport=0 0 455 830,clip,scale=0.4,angle=-90]{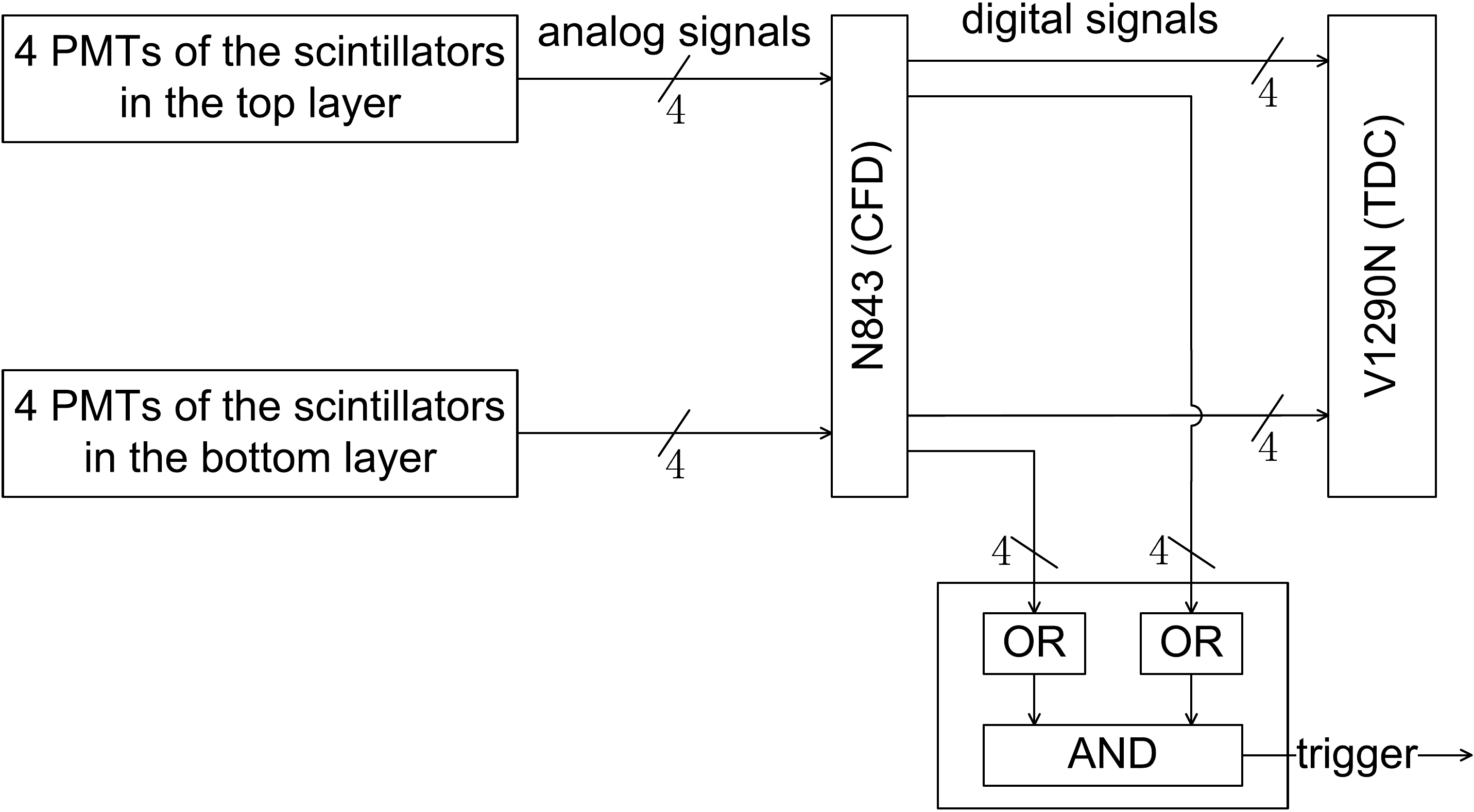}
  \end{minipage}
  \centering
  \caption{\small\sf (color online) Schematic layout of the detectors included in CoRaRS (left), and the processing of the PMT signals of the eight scintillator detectors (right).}
  \label{fig:detectors}
\end{figure*}

CORARS is $320.5cm$ high and $1.2\times1.2$ square meters in cross section as shown in Fig.~\ref{fig:detectors}. Both the top and bottom most layer consist of four scintillator detectors used for system trigger generation and time measurement, and a Thin Gap Chamber (TGC)~\citep{ATLAS1997ad} is placed in the inner side of each scintillator layer for position measurement of cosmic muons with a resolution of around $1cm$. Between the two TGC layers, the large space is for tested detectors. By design, maximum eight EDs can be inserted into the detector system to be tested simultaneously.

Each of the two TGCs is isosceles trapezium, with bases of $1.5m$ and $1.7m$, and height of $1.2m$. The 8 scintillator detectors are plastic ones in rectangular shape, with length of $1.2m$, width of $25cm$, and thickness of $7cm$, placed in parallel with small gaps of approximately $6.5cm$ between them to obtain a $1.2m\times1.2m$ areal coverage of CORARS.

\subsection{Electronics and DAQ} \label{subsec:principleAndTrigger}
When a cosmic muon enters CORARS, if the two scintillator layers are both fired, the event will be recorded. The time and spacial information of the cosmic muons entering CORARS are measured by scintillator layers and TGC layers, assuming the muon track is straight, which build the basis for the test of detectors.

The analog signals from the photo multiplier tubes (PMT) of scintillator detectors are discriminated in N843, which is a 16-channels Constant Fraction Discriminator (CFD) produced by CAEN company. If the amplitude of the input signal is larger than the threshold voltage set for the corresponding N843 channel, then a NIM signal is generated at $20\%$ of the leading edge of input signal. The NIM signal is then fed to CAEN V1290N, which is a 16-channel Time Digital Convertor (TDC), with LSB (Least Significant Bit) of $25ps$. And this NIM signal is also sent to logical modules CAEN V965 ("or" logic) and CAEN N455 ("and" logic) to participate in the trigger generation, as shown in Fig.~\ref{fig:detectors}. By setting the widths of the output NIM signals of N843 to $50ns$, a trigger is generated only when the time difference between the two scintillator layers' signals is smaller than $50ns$. This can effectively reduce the spurious triggers from noises.

Two-dimensional spacial information of the detected muon is given by TGC layers. There are 64 channels in one TGC layer, 32 of which are responsible for one dimension and the other 32 for the other dimension. The LVDS signals from the two TGCs' frond-end electronic boards are recognized and recorded by CAEN V1190A, a 128-channel TDC module. By then, the crossing points on the two TGC layers are recognized, and a straight muon track is reconstructed.

As the detectors to be tested, the time and charge of ED's signals are measured. Each signal is firstly fanned out into two identical signals. Then one is sent to CAEN V792N, which is a 16-channels charge-to-digital convertor (QDC), with LSB of $100fC$, and another is sent to N843 and then V1290N for time measurement.

A software has been developed using \texttt{c\#} to collect raw data from the VME modules, decode the data and then save it as ROOT files. This software also does some rudimentary data analysis on line to monitor the operational status of the system.


In order to eliminate the scintillating light propagation offset effect caused by the hitting position on the four scintillator detectors in the same layer, a timing calibration of the scintillator detectors was implemented. Each of the scintillator detectors was divided into many pixels of size $2.5cm\times2.5cm$, thanks to the spacial sensitivity of the two TGCs of the system. The mean time delay of every pixel relative to a global time reference was obtained. As a result of one-sided readout of the scintillator detector, the time delay increases along with the distance between the hit position and the PMT.  The width $\sigma_{delay}$ of the distribution of the time delay of each pixel serves as a figure-of-merit for the time resolution of that scintillator pixel. The $\sigma_{delay}$ of most of the pixels, $92\%$ to be precise, are smaller than $700ps$. And among those whose $\sigma_{delay}$ are larger than $700ps$, most are around the end side where the PMT is mounted. In order to keep a better time resolution, these pixels were not used.

\begin{figure*}[!hbt]
  \begin{minipage}{\columnwidth}
    \includegraphics[viewport=35 15 560 260,clip,scale=0.4]{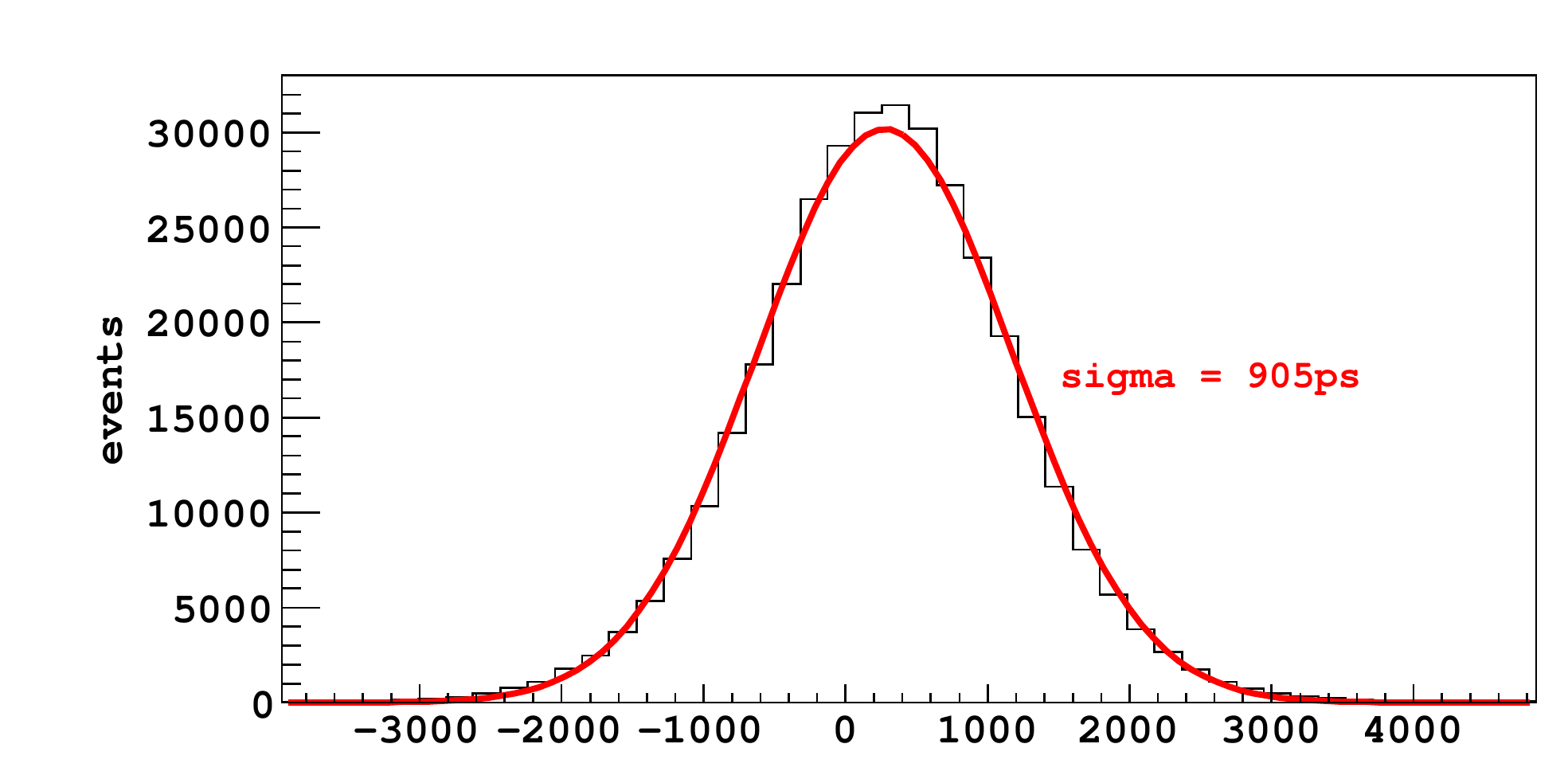}
    \centering
    \makebox{\scriptsize{($tof_{scin}-tof_{dis}$) (ps)}}
  \end{minipage}
  \hspace{3ex}
  \begin{minipage}{\columnwidth}
    \centering
    \includegraphics[viewport=30 10 565 250,clip,scale=0.4]{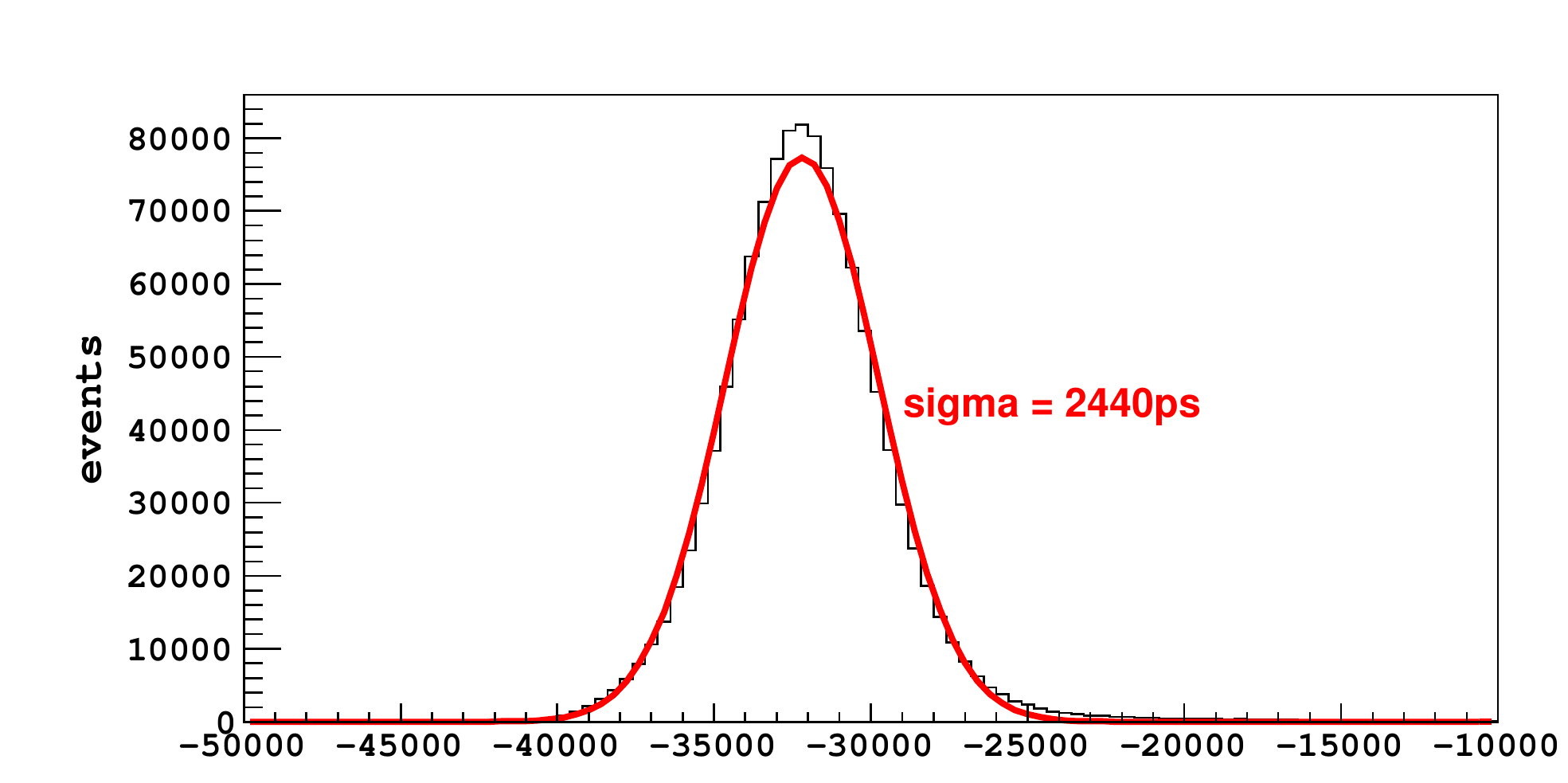}
    \centering
    \makebox{\scriptsize{($t_{ed}-t_{ed}^{reconstructed}$) (ps)}}
  \end{minipage}
  \hspace{3ex}
  \caption{\small\sf (color online) The distribution of the differences between the two time of flight obtained using two methods (left), and the distribution of the differences between the time given by ED and the reconstructed time by CORARS.(right).}
  \label{fig:sysSigma}
\end{figure*}

A test was done to find out the final time resolution of each layer of scintillator detectors after the calibration. For a muon that passes through the entire detector system, the time of flight of that muon can be obtained using
\begin{equation} \label{equ:tofScin}
tof_{scin}=(t_{btm}-t_{top}),
\end{equation}
where $t_{btm}$ is the time when the muon hits the bottom scintillator layer, and $t_{top}$ for the top scintillator layer. And a time of flight can also be obtained using
\begin{equation} \label{equ:tofDis}
tof_{dis}=d/v,
\end{equation}
where $d$ is the distance that the muon travels between the top and bottom scintillator layers, and $v$ is the velocity of the muon, which is assumed to be the same as the speed of light. Fig.~\ref{fig:sysSigma} shows the distribution of
\begin{equation}
\Delta tof=tof_{scin}-tof_{dis},
\end{equation}
and the width $\sigma_{\Delta tof}$ of this distribution is $905ps$. The $\sigma _{\Delta tof}$ is mainly due to the resolutions of the two scintillator layers. The resolutions of the two scintillator layers are supposed to be the same, so the time resolution of each scintillator layer after calibration is
\begin{equation} \label{equ:sigmaScin}
\sigma _{scin}=\sigma _{\Delta tof}/\sqrt 2\approx640ps.
\end{equation}

\section{Applications}
The performances of the first ED prototype was studied using this system, and the method and the results are presented.
\subsection{Time resolution and time resolution uniformity}
Using the time $t_{top}^{initial}$ given by the top scintillator layer, $t_{btm}^{initial}$ given by the bottom scintillator, and the calibration constants of the scintillator detectors, the corrected hitting time $t_{top}^{corrected}=t_{top}^{initial}-t_{top}^{delay}$ and $t_{btm}^{corrected}=t_{btm}^{initial}-t_{btm}^{delay}$ can be obtained. Then the time when the ED is hit can be reconstructed using the formula
\begin{equation}\label{equ:timeReconstructed}
t_{ed}^{reconstructed}=t_{top}^{corrected}+\frac{h}{H}\cdot(t_{btm}^{corrected}-t_{top}^{corrected}),
\end{equation}
where $H=320.5cm$ is the vertical distance between the top and the bottom scintillator layers, and $h=224.25cm$ is the vertical distance between the top scintillator layer and the layer where the prototype ED is placed. On the other hand, ED also gives the time $t_{ed}$ when it is hit by the muon. And the time resolution of this ED can be obtained from the width $\sigma_{\Delta}$ of the distribution of the time difference
\begin{equation} \label{equ:timeDiff}
\Delta=t_{ed}-t_{ed}^{reconstructed}.
\end{equation}
The histogram of this distribution is showed in Fig.~\ref{fig:sysSigma}, and a sigma $\sigma_{\Delta}=2440ps$ is obtained, where the contribution is from both the time resolutions of the ED and CORARS measurement. According to Eq.~\ref{equ:timeReconstructed}, the time resolution $\sigma_{t_{ed}^{reconstructed}}$ of reconstructed time $t_{ed}^{reconstructed}$ can be given by
\begin{equation} \label{equ:sigmaTimeRecon}
\sigma_{t_{ed}^{reconstructed}}=\frac{\sqrt{(H-h)^{2}+h^{2}}}{H}\cdot\sigma_{scin}\approx487ps,
\end{equation}
where $\sigma_{scin}=640ps$ (refer to Eq.~\ref{equ:sigmaScin}) is the time resolution of both $t_{top}^{corrected}$ and $t_{btm}^{corrected}$. Then a time resolution $\sigma_{ed}=2391ps$ of the ED is obtained using
\begin{equation} \label{equ:sigmaED}
\sigma_{ed}=\sqrt{(\sigma_{\Delta})^{2}-(\sigma_{ed}^{reconstructed})^{2}}\approx2391ps.
\end{equation}
This value is roughly consistent with the designed time resolution of $2ns$.



In order to study the time resolution uniformity of the ED, it is divided into $5cm\times5cm$ pixels. The mean value and the width $\sigma_{\Delta}$ of the distribution of $\Delta=t_{ed}-t_{ed}^{reconstructed}$ of each pixel was calculated, and the results are shown in Fig.~\ref{fig:timeResUniformity}.
\subsection{Photoyield and detection efficiency.}
\begin{figure*}[!hbt]
  \begin{minipage}{0.45\textwidth}
  \centering
  \includegraphics[viewport=30 10 550 270,clip,width=1\textwidth]{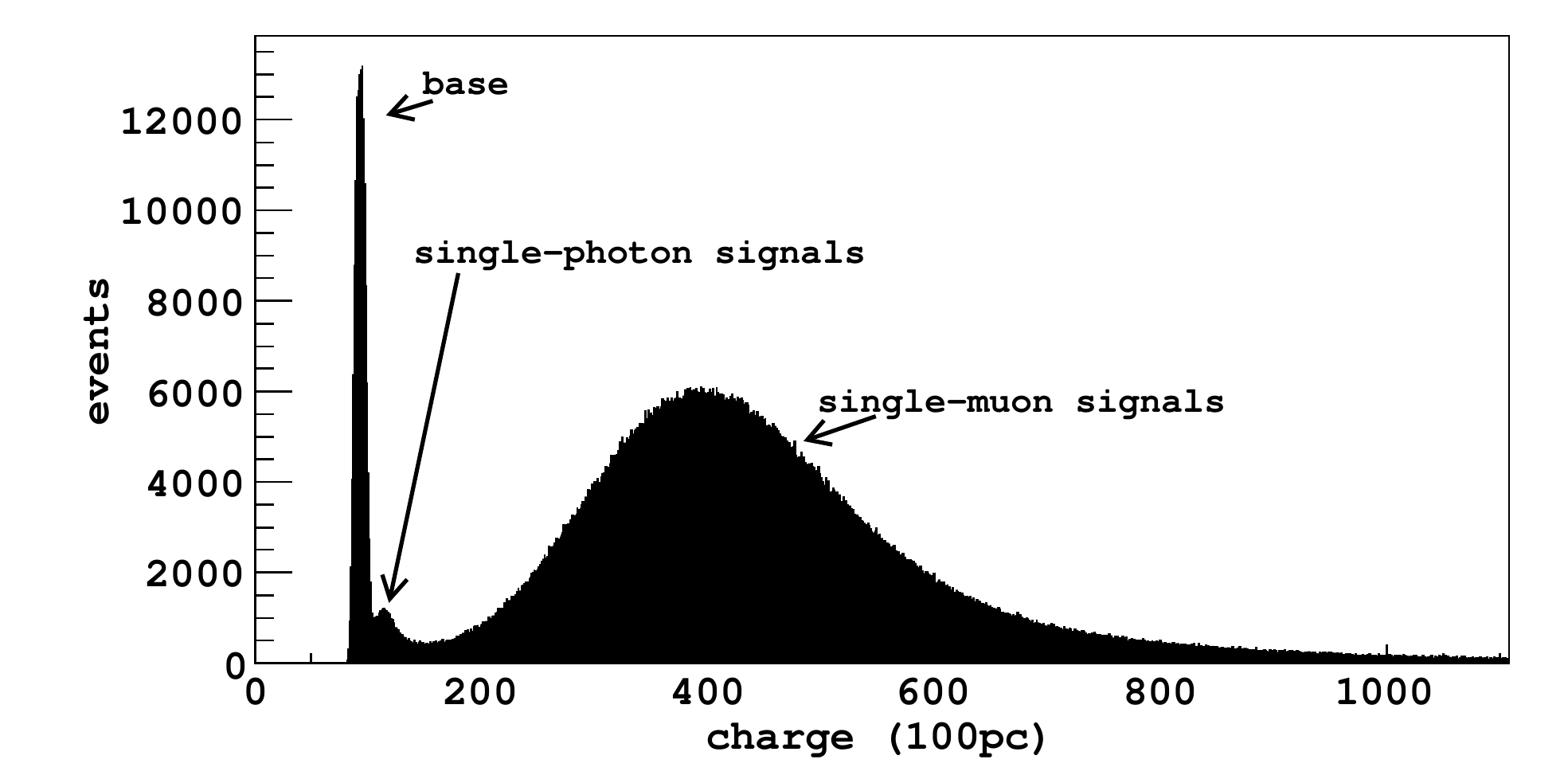}
  \end{minipage}
  \hspace{3ex}
  \begin{minipage}{0.45\textwidth}
  \centering
  \includegraphics[viewport=10 35 543 475,clip,width=1\textwidth]{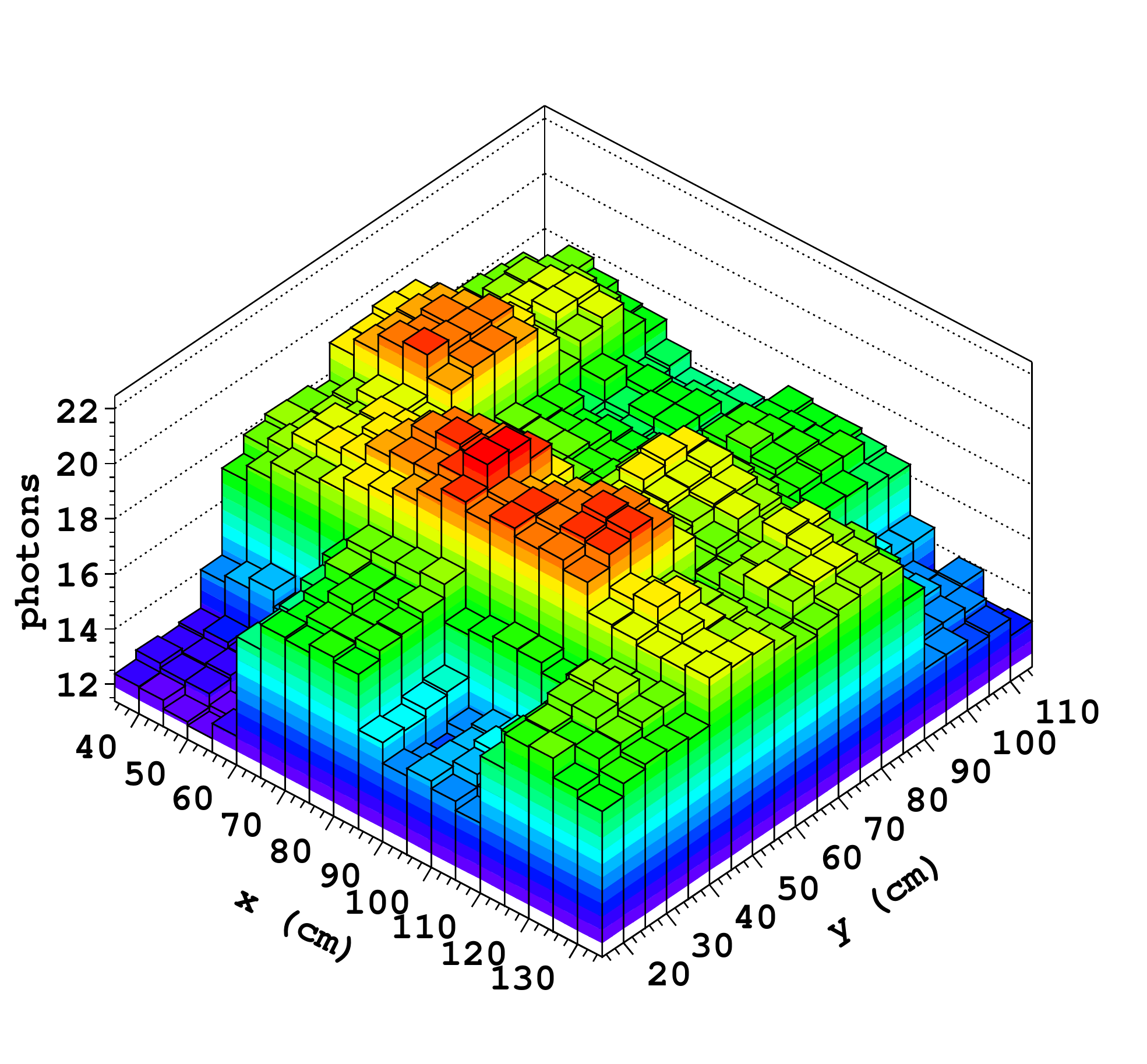}
  \end{minipage}
  \caption{\small\sf (color online) The charge spectrum of the ED signals (left), and the uniformity of the average photo-yield converted from the charge of ED signal (right).}
  \label{fig:q}
\end{figure*}
The charges of the signals from ED were measured. In the charge spectrum showed in the left part of Fig.~\ref{fig:q}, three peaks can be seen. The first peak is the electronic pedestal. The second peak is the charges of single-photon signals, and the last one is the charges of single-muon signals. The mean charge of pedestal is $9.3pc$, the mean charge of single-photon signals is $11.4pc$, and the single-muon signals $46.6pc$. It can be calculated that $17.8$ photon-electrons on average are collected by PMT when one cosmic muon passes through ED, i.e., the average photoyield is $17.8 photon/muon$. The photoyield of each $5cm\times5cm$ pixel of ED was calculated and is shown in the right part of Fig.~\ref{fig:q}.

\begin{figure*}[!hbt]
  \begin{minipage}{0.45\textwidth}
  \centering
  \includegraphics[viewport=55 80 520 430,clip,scale=0.45]{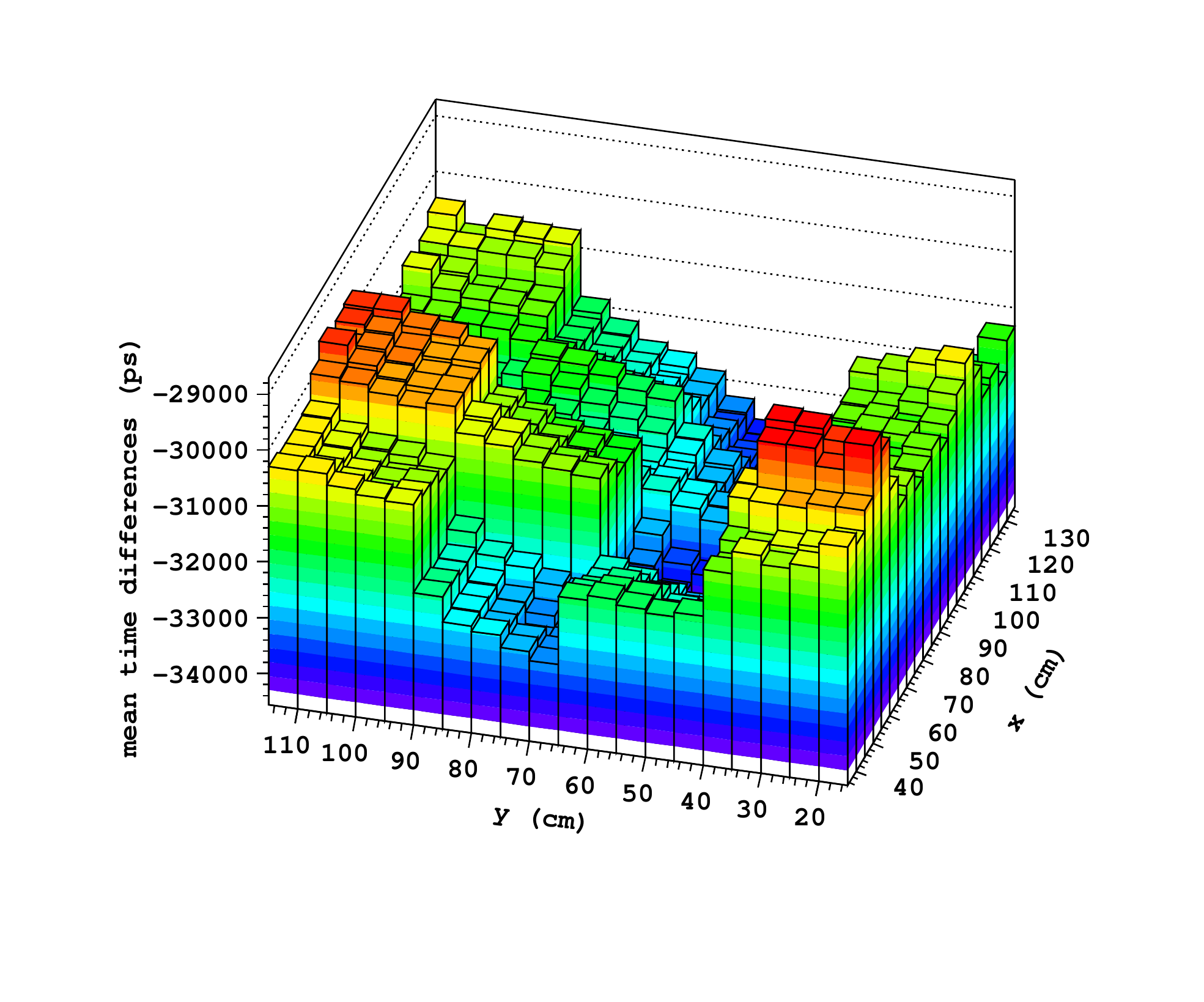}
  \end{minipage}
  \hspace{3ex}
  \begin{minipage}{0.45\textwidth}
  \centering
  \includegraphics[viewport=15 60 560 500,clip,scale=0.4]{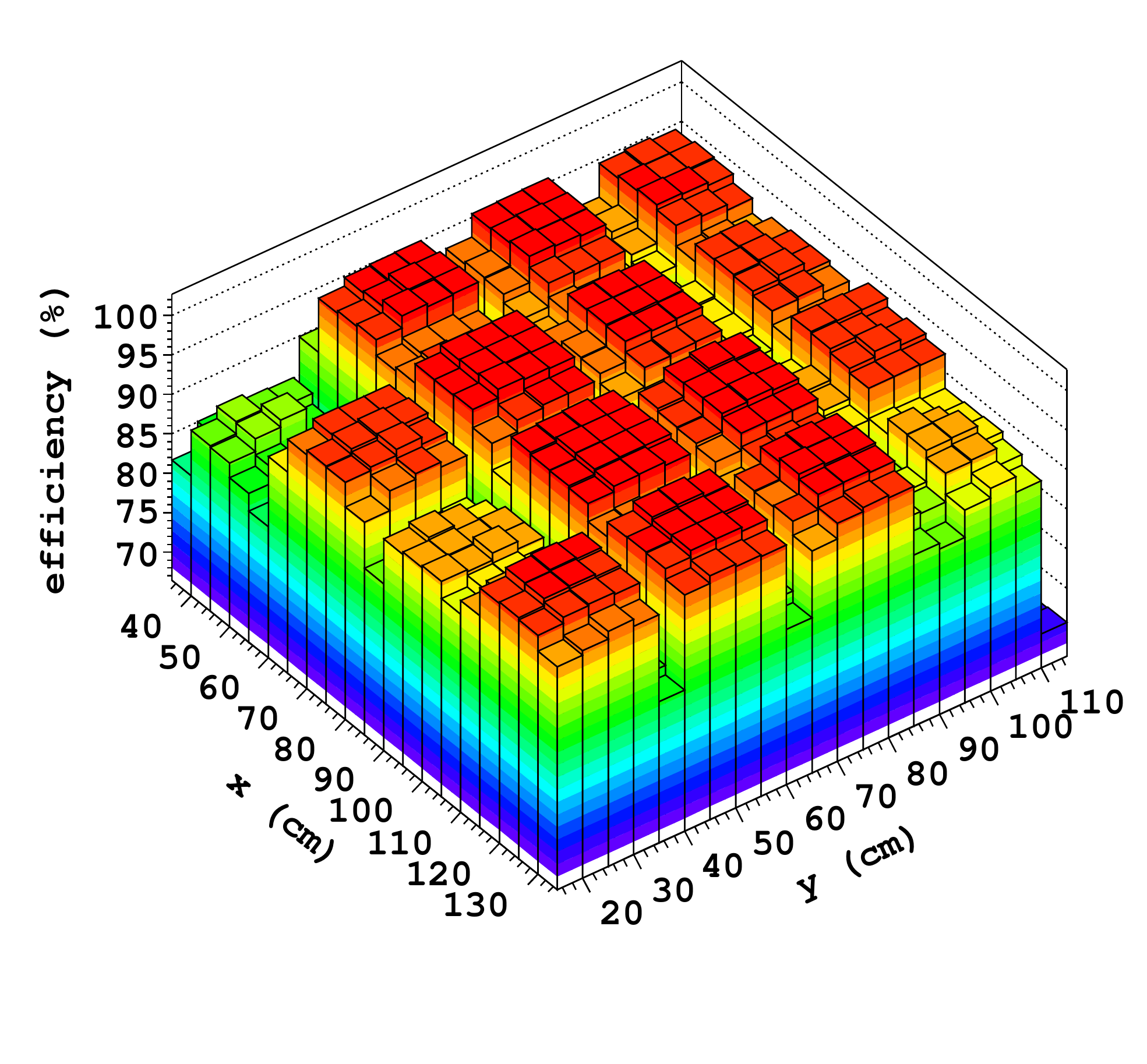}
  \end{minipage}
  \caption{\small\sf (color online) The uniformity of the time resolution (left) and the the detection efficiency (right) of each $5cm\times5cm$ pixel of ED. }
  \label{fig:timeResUniformity}
\end{figure*}

The detection efficiency of the ED was studied as well. When a cosmic muon is confirmed by CORARS to have hit the ED, the signal from ED is checked. If ED signal presents, this muon is considered as being detected by ED, otherwise it is considered as being missed by ED. The radio of the number of events being detected by ED to the number of total events is the detection efficiency of the ED. A average detection efficiency of about $93.7\%$ was obtained for the entire ED. The detection efficiency of each $5cm\times5cm$ pixel was also calculated, and the result is showed in Fig.~\ref{fig:timeResUniformity}. This ``lower'' efficiency is partially due to threshold definition of the electronic system, and can be adjusted later.
\section{Summary}
A testing system has been built for general detector testing, which has good time and spacial resolution. The first ED prototype for LHAASO is tested, showing it has reached the designed level. A time resolution $2.4ns$ of ED is obtained. And about 17.8 photon-electrons are collected averagely when a cosmic muon passes through it. The mean detection efficiency of the entire ED is about $93.7\%$. \\








\end{multicols}

\clearpage
\end{CJK*}

\end{document}